\documentclass[sts]{imsart}

\RequirePackage[OT1]{fontenc}
\usepackage{amsthm,amsmath,natbib, amssymb}
\usepackage{graphicx, multicol, array, algorithm}
\RequirePackage[colorlinks,citecolor=blue,urlcolor=blue]{hyperref}

\startlocaldefs
\numberwithin{equation}{section}
\theoremstyle{plain}

\def\dR{\mathbb{R}}

\def\simind{\stackrel{{\tiny \mbox{ind.}}}{\sim}}
\def\cmpMu{\text{CMP}_{\mu}}

\def\bI{\boldsymbol{I}}
\def\bzero{\boldsymbol{0}}

\def\dR{\mathbb{R}}

\def\btheta{\boldsymbol{\theta}}

\def\bX{\boldsymbol{X}}
\def\by{\boldsymbol{y}}

\def\bbeta{\boldsymbol{\beta}}

\endlocaldefs

\begin{document}

\begin{frontmatter}
\title{\huge Bayesian generalized linear model for over and under dispersed counts}
\runtitle{Bayesian GLM for dispersed counts}

\begin{aug}
\author{\fnms{Alan} \snm{Huang$^1$}\ead[label=e1]{alan.huang@uq.edu.au}},
\and
\author{\fnms{Andy Sang Il} \snm{Kim$^2$}\ead[label=e2]{andykim86@gmail.com}}

\runauthor{A. Huang \& A. S. I. Kim}

\affiliation{$^1$University of Queensland, Nanyang Technological University, and $^2$University of Technology Sydney}

\address{School of Mathematics and Physics, University of Queensland, St Lucia, QLD, Australia 4072 \printead{e1}.}

\address{School of Mathematical and Physical Sciences, University of Technology Sydney, Ultimo, NSW, Australia 2007 \printead{e2}.}
\end{aug}

\begin{abstract}
Bayesian models that can handle both over and under dispersed counts are %particularly 
rare in the literature, perhaps because full probability distributions for dispersed counts are rather difficult to construct. This note takes a first look at Bayesian Conway-Maxwell-Poisson generalized linear models that can handle both over and under dispersion yet retain the parsimony and interpretability of classical count regression models. The focus is on providing an explicit demonstration of Bayesian regression inferences for dispersed counts via a Metropolis-Hastings algorithm. We illustrate the approach on two data analysis examples and demonstrate some favourable frequentist properties via a simulation study.
\end{abstract}


\end{frontmatter}

\section{Introduction}
Count data often exhibit dispersion relative to a classical Poisson model. For overdispersed counts, where the conditional variance is larger than the conditional mean, models based on the negative binomial, Poisson inverse-Gaussian and other Poisson rate mixture distributions have been well-covered in the literature from both frequentist and Bayesian points of view \citep[e.g.,][]{Wilmot1987,MN1989,GH2007}. For underdispersed counts, where the conditional variance is smaller than the conditional mean, the options are far more limited due to the difficulty in constructing full probability distributions that can handle underdispersion. A recent survey of a handful of existing models can be found in \citet{SM2017}. The lack of options is particularly detrimental from a Bayesian point of view because a full probability distribution for the data is required for the application of Bayes' theorem to determine the posterior distribution and subsequent inferences. 

This note takes a first step towards filling this gap in the literature by presenting a Bayesian framework for the regression modelling of counts that can (i) handle both over and under dispersion, and (ii) retain the same level of parsimony and interpretability as classical counts models. We do this by building on a recent reparametrisation of the Conway-Maxwell-Poisson distribution (CMP) that allows the mean to be modelled directly \citep{Huang2017}, so that simple and interpretable models, such as a log-linear model, can be constructed for both over and under dispersed counts. This places underdispersed counts on the same footing as equidispersed and overdispersed counts which are readily handled by the familiar Poisson and negative binomial models, respectively.

More specifically, the mean-parametrized CMP distribution with mean $\mu > 0$ and dispersion $\nu \ge 0$ is characterized by the probability mass function
\begin{equation}
P(Y=y \vert \mu,\nu) = \frac{\lambda(\mu, \nu)^y}{(y!)^\nu} \frac{1}{Z(\lambda(\mu, \nu), \nu)} \ , \quad y = 0,1,2,\ldots \ ,
\label{eq:cmpExp}
\end{equation}
where the rate $\lambda(\mu, \nu)$ is a function of $\mu$ and $\nu$ given by the solution to
$$
0 = \sum_{y=0}^{\infty} (y-\mu)\frac{\lambda^{y}}{(y!)^{\nu}} \ ,
$$
and $Z(\lambda, \nu) = \sum_{y=0}^\infty \lambda^y/(y!)^\nu $ is a normalizing function. It is easy to show that $\nu < 1$ implies overdispersion and $\nu > 1$ implies underdispersion relative to a Poisson distribution of the same mean. When $\nu =1$, the distribution coincides with the Poisson distribution. We write CMP$_\mu(\mu, \nu)$ for the mean-parametrized CMP distribution (\ref{eq:cmpExp}) to distinguish it from the standard CMP distribution of \cite{SMKBB2005}.

CMP$_\mu$ distributions are particularly useful for modelling dispersed counts because they retain all the attractive properties of standard CMP distributions whilst being able to model the mean directly \citep{Huang2017,Andrianatos2017}. They form two-parameter exponential families, and for fixed or given values of the dispersion they become one-parameter exponential families, making them immediately adaptable to regression modelling via the generalized linear model \citep[GLM,][]{MN1989} framework. They also form a continuous bridge between other classical models, passing through the geometric and Poisson distributions as special cases, with the Bernoulli distribution as a limiting case. Moreover, unlike the generalized Poisson \citep{CF1992}, quasi-Poisson \citep{Wedderburn1974}, and the recently-proposed Extended Poisson-Tweedie \citep{BJKHD2018} models, CMP$_\mu$ distributions always correspond to full probability models for any choice of model parameters, covering both over and under dispersion. This makes them particularly suitable for Bayesian modelling and inferences.

From a Bayesian point of view, \cite{KSMBB2006} explored the use of conjugate priors for the standard CMP distribution in the independent and identically distributed case. While this can be easily extended to mean-parametrized CMP distributions (see Online Supplement), there are three practical limitations of using the conjugate prior. First, the subsequent posterior is known only up to a normalizing constant that has no closed-form, so computing it requires either numerical integration, Markov Chain Monte Carlo (MCMC) sampling, or some other approximation method. There is therefore no practical advantage of using the conjugate prior over any other prior.
Second, appropriate specification of the hyperparameters can be rather opaque due to the unfamiliar form of the conjugate prior. \cite{KSMBB2006} offer an applet to translate prior information into an appropriate elucidation of hyperparameters, but a simpler and more interpretable prior can mitigate this issue altogether. 
Finally, the conjugacy property only holds for intercept-only models, and the extension to the regression case in which the mean of each observation may depend on a set of covariates is not at all immediate. This is in fact true of most Bayesian generalized linear models -- for example, while the Gamma distribution is conjugate for the Poisson distribution for intercept-only models, there is no extension to the Poisson regression case \citep[][page 173]{Hoff2009}. Instead, multivariate normal priors are placed on the regression coefficients for interpretability and parsimony (as implemented in the popular \texttt{MCMCpack} package in \texttt{R} by \citet{MQP2011}, for example).

To the best of our knowledge, this note is the first to consider a Bayesian framework for the regression modelling of both over and under dispersed counts that circumvents these limitations, yet retains the parsimony and interpretability of familiar count models such as the log-linear Poisson and negative binomial models. This work is part of a larger ongoing project that looks at efficient Bayesian inferences for dispersed counts in hierarchical models.

\section{Bayesian CMP generalized linear model for dispersed counts}
Suppose we have independent observation pairs $(y_1,\bX_1),(y_2, \bX_2)  \dots$, where each $y_i$ is a count response and each $\bX_i\in \dR^p$ is a corresponding set of covariates. A CMP$_\mu$ generalized linear model for the data that can handle both over and under dispersion can be specified via
\begin{equation}
y_i \vert \bX_i \stackrel{\rm ind.}{\sim} \cmpMu \left(\mu(\bX_i^\top \bbeta), \nu \right) \ , \quad i=1,2,\ldots \ ,
\label{eq:meanCmpReg}
\end{equation}
where $\bbeta \in \dR^p$ is a vector of regression coefficients, $\nu \ge 0$ is a dispersion parameter, and (in a slight abuse of notation) $\mu(\cdot)$ is a user-specified mean-model or inverse-link function. For this note, we focus on the popular log-linear model,
\begin{eqnarray*} 
\begin{array}{c}
E(y \vert\bX) = \mu(\bX^\top\bbeta) = \exp(\bX^\top\bbeta),
\end{array}
\label{eq:cmpMeanFunction}
\end{eqnarray*}
so that each component of $\bbeta$ can be interpreted as the expected change in the mean response (on the log scale) for a unit increase in the corresponding component of $\bX$. Of course, other link functions can be considered, as with any GLM. Indeed, the key advantage of the CMP$_\mu$ distribution is that is directly parametrized via the mean so that simple, easily-interpretable mean-models can be considered. In contrast, standard CMP distributions \citep{SMKBB2005, LGG2008, SS2010} and its variants \citep[e.g.,][]{GG2008} model either a latent rate parameter or some power transformation of it, which cannot be interpreted as the mean.

For a Bayesian model specification, we need a prior distribution on the model parameters $\bbeta$ and $\nu$. In this note, we focus on easily interpretable prior specifications, such as
\begin{eqnarray} 
\begin{array}{ccc}
\bbeta \sim \mbox{N}(\mu_{\bbeta}, \Sigma_{\bbeta}) \quad \text{and} & \nu \sim \text{Log-Normal} (\mu_\nu, \sigma^2_\nu) \ ,
\end{array}
\label{eq:meanCmpRegBayes}
\end{eqnarray}
so that the hyperparameters $(\mu_{\bbeta},\Sigma_{\bbeta})$ and $(\mu_\nu, \sigma_\nu^2)$ have clear interpretations as prior means and variances. This makes it easy to translate prior beliefs about the data-generating process into sensible specification of hyperparameter values. In contrast, the conjugate prior of \cite{KSMBB2006} has no clear interpretation, making elucidation of hyperparameter values rather opaque. Of course, the prior distribution in (\ref{eq:meanCmpRegBayes}) can be replaced with any user-specified prior, with the proposed method in this note being applicable for any prior specification. Note that taking the prior variances in (\ref{eq:meanCmpRegBayes}) to be arbitrarily large leads to improper flat priors, $p(\bbeta), p(\nu) \propto 1$, which we consider in Section \ref{sec:sensitivity} and in the Online Supplement.

When the normal and log-normal priors (\ref{eq:meanCmpRegBayes}) are used, the joint posterior of the parameters $\bbeta$ and $\nu$ given the observed data $\by \equiv (y_1, \dots, y_n)$ has the form,
\begin{eqnarray}
\label{eq:postMeanCmpReg}
p(\bbeta,\nu\vert\by,\bX) 
&\propto&
\prod_{i=1}^{n} \lambda \left(\exp(\bX_i^\top \bbeta),\nu \right) ^{y_i} Z\left(\lambda(\exp(\bX_i^\top \bbeta),\nu),\nu \right)^{-1} y_i^{-\nu}
\\ && \nonumber
\times \exp\left[-\frac{1}{2}(\bbeta - \mu_{\bbeta})^\top \Sigma_{\bbeta}^{-1} (\bbeta - \mu_{\bbeta}) \right]
\\ && \nonumber
\times \nu^{-1} \exp \left[ - \frac{(\log \nu - \mu_\nu)^2}{2 \sigma_\nu^2} \right] \ ,
\end{eqnarray}
which is known up to a normalizing constant. Direct computation of this posterior distribution is not possible, but the explicit form of the density kernel in (\ref{eq:postMeanCmpReg}) means that a simple Metropolis-Hastings (MH) algorithm can be directly used. We illustrate this in Section 3.

A reviewer pointed out that a key advantage of the Bayesian approach over the frequentist approach of \citet{Huang2017} is that posterior predictive distributions can be obtained for a new observation $\tilde y$ corresponding to some covariate $\tilde \bX$, given the observed values, via
$$
p(\tilde y  \vert  \tilde \bX, \by) = \int p(\tilde y \vert \bbeta, \nu, \tilde \bX) \ p(\bbeta, \nu \vert \by, \bX) \ d(\bbeta, \nu) \ .
$$
This can be estimated via Monte Carlo averaging of the likelihood $p(\tilde y \vert \bbeta, \nu, \tilde \bX)$ evaluated at draws $(\bbeta,\nu)$ from the posterior distribution (\ref{eq:postMeanCmpReg}) obtained from the proposed MH algorithm.

\section{Bayesian inferences for dispersed counts}
From a Bayesian point of view we are interested in evaluating posterior distributions of the form (\ref{eq:postMeanCmpReg}) and making inferences on model parameters $(\bbeta,\nu)$ via computing posterior means and credible intervals. Recent work by \cite{CENN2017} presents an efficient approach to Bayesian inference for CMP models via a rejection sampling based method called the ``exchange algorithm", which is applicable to situations where the likelihood function can be computed only up to a normalising constant. This novel algorithm does not require computation of normalizing constants for the Metropolis-Hastings acceptance ratio by incorporating an auxiliary data $\by^*$ generated from the sampling model $p(\by^*\vert\btheta^*)$ where $\btheta^*$ is drawn from the proposed distribution of the model parameters $\btheta$. This approach allows us to cancel out all intractable normalising constants. Detailed explanation of the algorithm and its comparison with the standard Metropolis-Hastings algorithm is presented in Section 3.2 of \cite{CENN2017}. \cite{Benson2017} also develop a new, faster CMP rejection sampler by building two tractable enveloping bounds. They then link the sampler with the reciprocal normalising constant, which allows the intractable likelihood function to be estimated without bias. While these approaches work nicely for regression models that are based on the standard CMP of \cite{SMKBB2005} and the reparameterized CMP by \cite{GG2008}, they have not (yet) been adapted for the mean-parametrized $\cmpMu$ distribution -- this has been flagged for immediate future research.

Instead, we present a direct Metropolis-Hastings (MH) algorithm that is particularly straightforward to implement. There is also no restriction on the types of proposal densities that can be used, and we can choose to alternate updates for each parameter component which we have found to improve convergence speed of the subsequent Markov Chain. 

For simplicity of illustration, we consider multivariate normal proposals for $\bbeta$ in our MH algorithm, with mean of each proposal given by the current value $\bbeta_0$  and some covariance matrix $S_{\bbeta}$. That is, we sample proposals $\bbeta_1 \sim \mbox{N}(\bbeta_0, S_{\bbeta})$. Due to the symmetry of normal distributions, the acceptance probability of this proposal is simply the ratio of the likelihoods, 
\begin{equation}
\label{eq:acceptRatioMeanCmpRegBeta} 
a_{\beta} = \frac{p(\bbeta_1, \nu_0| \by, \bX)}{p(\bbeta_0, \nu_0| \by, \bX)} \ ,
\end{equation}
where $\nu_0$ is the current value of $\nu$. Similarly, we can consider exponential proposals for $\nu$ with mean of each proposal given by the current value $\nu_0$. That is, we sample proposals $\nu_1 \sim \mbox{Exp}(1/\nu_0)$ with density $p(\nu_1) = \nu_0^{-1} \exp(-\nu_1/\nu_0)$. The corresponding acceptance probability is therefore given by 
\begin{eqnarray}
a_\nu &=&
\frac{p(\bbeta_0,\nu_1\vert \by, \bX)}{p(\bbeta_0,\nu_0 \vert \by, \bX)} \frac{\nu_1}{\nu_0} \exp\left( \frac{\nu_1}{\nu_0} -\frac{\nu_0}{\nu_1} \right)  \ ,
\label{eq:acceptRatioMeanCmpRegNu}
\end{eqnarray}
where  $\bbeta_0$ is the current value of $\bbeta$. Of course, other proposal distributions can also be used. However, we find the  exponential proposal distribution to be particularly simple to work with as there is no secondary ``variance"-type parameter to select. Indeed, the normal--exponential pair of proposal distributions led to excellent mixing times in the all examples that we looked at.

The MH MCMC algorithm obtained via alternating these proposals is summarized in Algorithm \ref{alg:compRegBayes} below. We readily admit that this is by no means the most efficient algorithm possible -- indeed, this note is part of a larger study on efficient Bayesian inferences for dispersed counts in hierarchical models. However, we believe that this is the first framework that places the analysis of underdispersed counts on the same footing as classical Poisson and negative binomial regression models for equidispersed and overdispersed counts, respectively. The algorithm is implemented as a simple plug-in to the \texttt{mpcmp} package \citep{FAWA2019} in \texttt{R}, and can be obtained via emailing the corresponding author.
%%%%%%%%%%%%%%%%%%%%%%%%%%%%%%%%%%%%%%%%%%%%%%%%%%%%%%
\begin{algorithm}
\begin{center}
\begin{minipage}[t]{150mm}
%\hrule
%
%\vskip2mm\noindent
%
Inputs: data $\bX, \by$
\vskip2mm\noindent%
Initialize: $\bbeta$ and $\nu$ at some $\bbeta_0$ and $\nu_0$ respectively.
\vskip2mm\noindent
Cycle through: 
\begin{itemize}
\item[1] \quad Draw a sample $\bbeta_1$ from $\bbeta_1 \sim N (\bbeta_0,S_{\bbeta})$
\item[2] \quad Accept $\bbeta_1 $ with probability $\text{min}\left(1,a_{\bbeta} \right)$ and update $\bbeta_0 \leftarrow \bbeta_1$
\item[3] \quad  Else reject $\bbeta_1$ and keep $\bbeta_0$
\item[4] \quad  Draw a sample $\nu_1$  from  $\nu_1 \sim \text{Exp} (1/\nu_0)$
\item[5] \quad Accept $\nu_1 $ with probability $\text{min}\left(1,a_\nu \right)$ and update $\nu_0 \leftarrow \nu_1$
\item[6] \quad  Else reject $\nu_1$ and keep $\nu_0$.
\end{itemize}
\vskip1mm
until $N$ MCMC samples are generated.   
%
%\hrule
\end{minipage}
%\hrule
\end{center}
\caption{\it An alternating MH MCMC algorithm for sampling from the posterior density $p(\bbeta, \nu \vert \by, \bX)$ in the $\cmpMu$ generalized linear model \eqref{eq:meanCmpReg}--\eqref{eq:meanCmpRegBayes}.}
\label{alg:compRegBayes} 
%\hrule
\end{algorithm}

\section{Numerical examples}
The proposed framework is illustrated on two data analysis examples, one exhibiting overdipsersion and the other underdispersion. These examples are accompanied by a simulation study that demonstrates some favourable frequentist properties of the approach. All examples and simulations were carried out in {\rm R} 3.4.1 (R Development Core Team, 2017) on an iMac desktop computer with an i7 3.4GHz Intel Core CPU and 16.0GB RAM. 

To implement Algorithm \ref{alg:compRegBayes} in practice, we first obtain the MLE of the model parameters for two purposes. First, the MLE estimates $\hat \bbeta_{\rm MLE}$ and $\hat \nu_{\rm MLE}$ are used to initialize the starting values $\bbeta_0$ and $\nu_0$ in the MCMC sampler. 
This replaces the burn-in period. 
Second, the estimated variance-covariance matrix $\hat \Sigma_{\rm MLE}$ of $\hat \bbeta_{\rm MLE}$ is used for the variance matrix $S_{\bbeta}$ in the proposal distribution for $\bbeta$. For each real or simulated example, $1000$ MCMC samples were generated with a thinning factor of 10. Thinning was used to reduce the autocorrelation in the samples.

\subsection{Overdispersed class attendance data}
\begin{figure}[]
\centering
\includegraphics[width= 0.8 \textwidth]{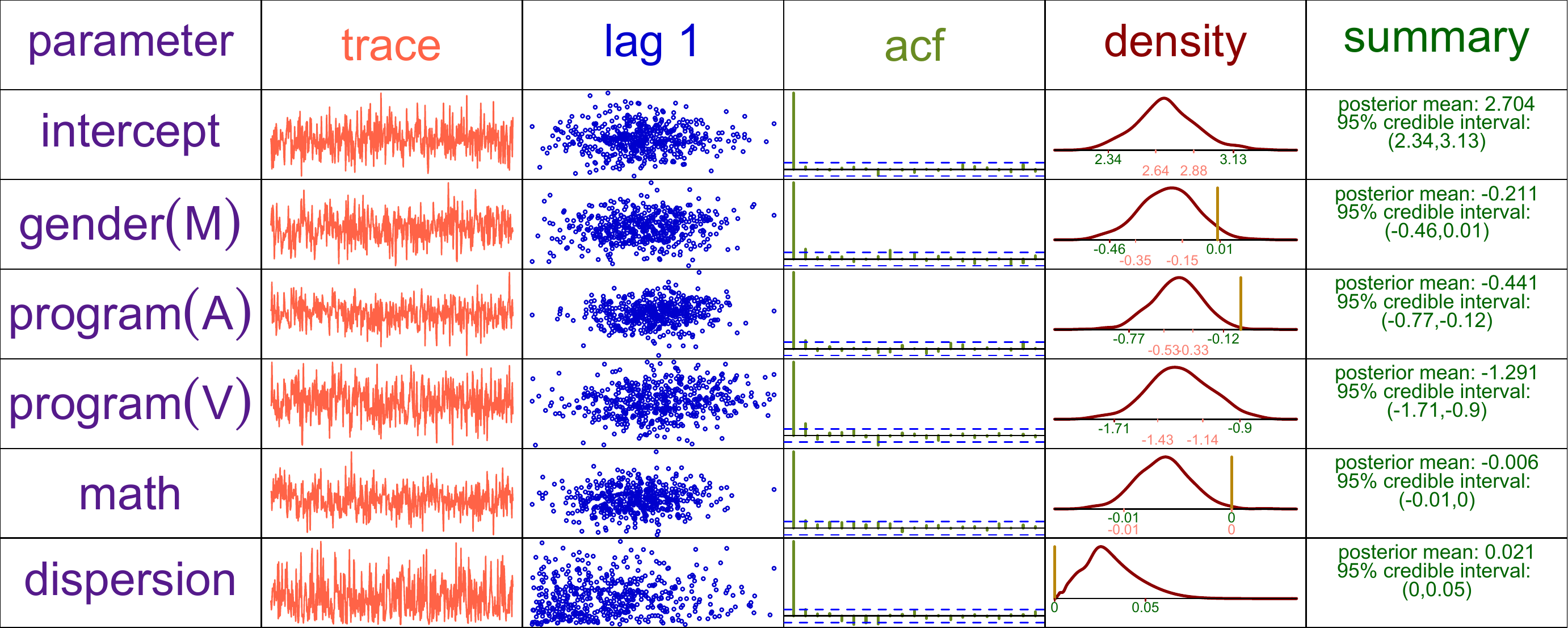}
\caption{Summary of MH MCMC-based inference for parameters in the $\cmpMu$ log-linear Bayesian regression model \eqref{eq:meanCmpRegBayes} fitted to the class attendance data. The columns are: (1) parameter name, (2) trace plot of the  MCMC sample, (3) plot of MCMC sample against its lag 1 sample, (4) sample autocorrelation functions, (5) kernel density estimate of posterior density function, (6) numerical summaries of posterior density function. }
\label{fig:daysResult}
\end{figure}

The $\cmpMu$ GLM \eqref{eq:meanCmpReg}--\eqref{eq:meanCmpRegBayes} is used to analyze the attendance dataset from 
%\textit{http://www.ats.ucla.edu/stat/stata/dae/nb_data.dta} 
\url{http://www.ats.ucla.edu/stat/stata/dae/nb_data.dta} 
which contains the total number of days absent in an academic year for each of 314 students sampled from two urban high schools. Explicitly, our model for the data is the log-linear CMP$_\mu$ regression model,
\begin{eqnarray}
\begin{array}{rcl}
y_i \vert \bbeta,\nu & \simind & \cmpMu\left(\exp\left\{\beta_0 + \beta_1\text{Female}_i + \beta_2\text{Academic}_i \right.\right. \\
& & \left.\left.
+ \beta_3\text{Vocational}_i + \beta_4\text{Math}_i \right\},\nu \right) \ ,
\end{array}
\label{eq:classAttModel}
\end{eqnarray}
where for each student $i$, $y_i$ is the total number of days absent, $\text{Female}_i$ is an indicator for being female, $\text{Academic}_i $ is an indicator for being in the academic program, $\text{Vocational}_i $ is an indicator for being in the vocational program, and $\text{math}_i $ is the mathematics score. For our priors we consider
\begin{eqnarray}
\begin{array}{cc}
\bbeta \sim N(\bzero,10^5 \bI), & \quad \nu \sim \text{log-Normal}(0,10^5) \ ,
\end{array}
\label{eq:classAttPrior}
\end{eqnarray}
corresponding to vague {\it a priori} beliefs about the model parameters. Algorithm \ref{alg:compRegBayes} was used to obtain a sequence of random draws from the posterior distribution arising from model \eqref{eq:classAttModel}--\eqref{eq:classAttPrior}. Graphical summaries of the MCMC output and subsequent Bayesian inferences for the model parameters using the visualisation tool from \citet{MW2010} are presented in Figure \ref{fig:daysResult}. 

% % 

We see from column 2 of Figure \ref{fig:daysResult} that even with no burn-in the trace plots indicate that the Markov chain exhibited good mixing. There is also negligible correlation with the lagged-one sample (column 3), as well as minimal autocorrelation across all other lags (column 4). Column 5 plots the kernel estimates of the marginal posterior densities for all model parameters. In these plots, we also overlay the endpoints of $95\%$ credible intervals (marked in maroon) as well as those from a standard Poisson log-linear model (marked in pink). The last column provides a numerical summary of the MCMC estimates with the posterior means and the endpoints of the 95\% credible intervals. 

The use of the simple log-linear mean-model allows the posterior estimates to be easily interpreted. For example, the posterior mean of $\beta_1$ estimates that female students miss $\exp(0.211) = 1.23$ times as many days of school compared to male students, with a 95\% credible interval between $\exp(-0.01)=0.99$ and $\exp(0.46)=1.58$ times. Analogously, students in the academic and vocational programs are each expected to miss an estimated $\exp(0.441)=1.55$ and $\exp(1.291)=3.64$ times as many days, respectively, than students in the General (baseline) program. Comparing the credible intervals of the $\cmpMu$ model with that of the standard Poisson model, we note that the Poisson model has not accounted for the conditional overdispersion in the data, as evidenced by the narrower intervals across all regression coefficients. Finally, the estimated dispersion parameter has posterior mean of $0.021$ with 95\% credible interval of $(0,0.05)$, confirming strong conditional overdispersion in the data. Each MCMC update took approximately 0.2 seconds on average to run. 

\subsection{Underdispersed takeover bids data}
\begin{figure}[]
\centering
\includegraphics[width= 0.8 \textwidth]{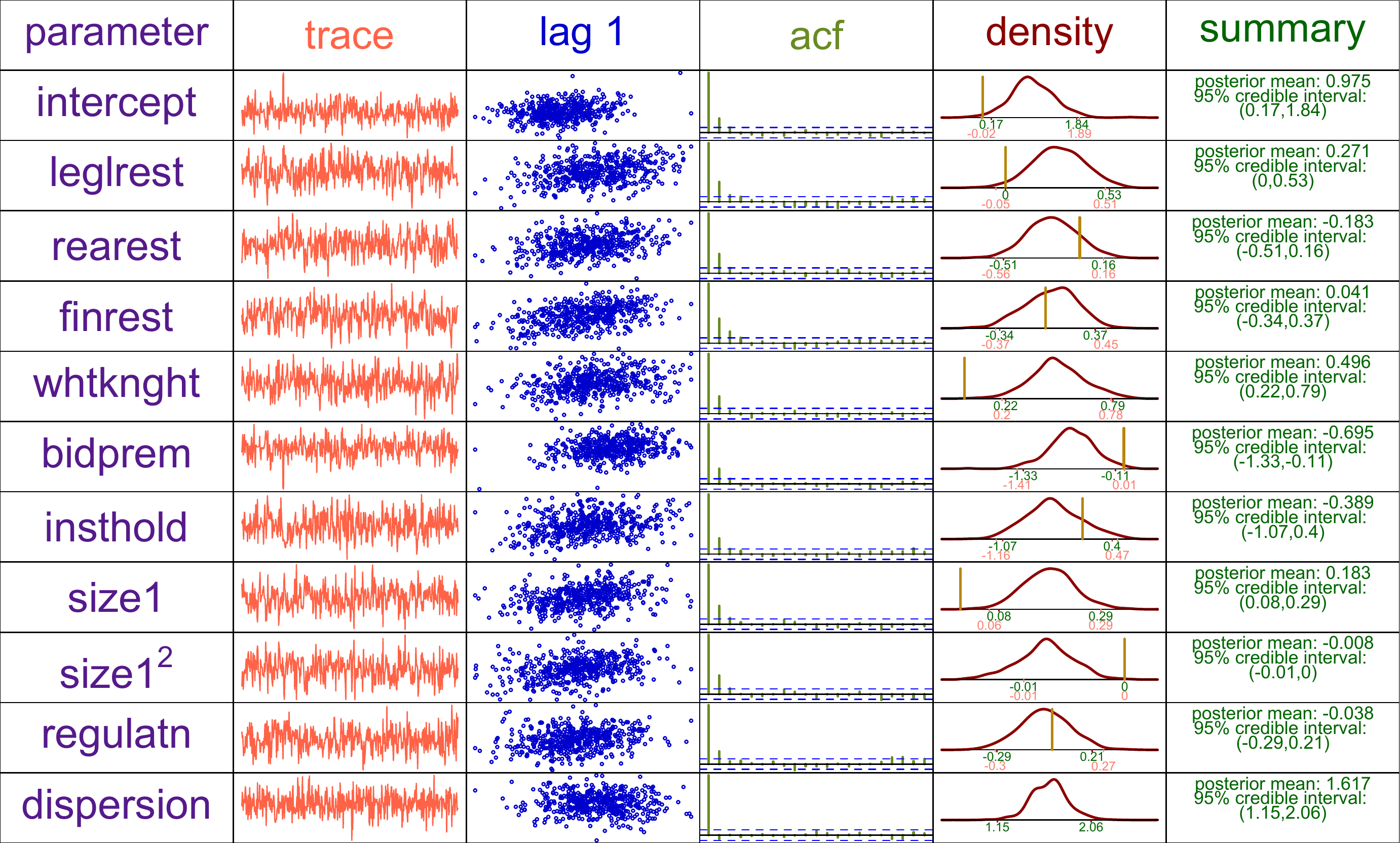}
\caption{Summary of MH MCMC-based inference for parameters in the $\cmpMu$ log-linear Bayesian regression model \eqref{eq:meanCmpRegBayes} fitted to the takeover bids data.}
\label{fig:bidResult}
\end{figure}

The $\cmpMu$ GLM \eqref{eq:meanCmpReg}--\eqref{eq:meanCmpRegBayes} is also used to analyze the takeover bids data from \cite{SS2013}. For this dataset we consider a model of the form, 
\begin{eqnarray}
\begin{array}{rcl}
\text{bids}_i \vert \bbeta,\nu & \simind & \cmpMu\left[\exp\left\{\beta_0 + \beta_1\text{leglrest}_i + \beta_2\text{rearest}_i  + \beta_3\text{finrest}_i  \right.\right. \\
& & \left.\left. + \beta_4\text{whtknght}_i + \beta_5\text{bidprem}_i + \beta_6\text{insthold}_i + \beta_7\text{size}_i  \right.\right. \\
& & \left.\left. + \beta_8\text{size}^2_i + \beta_9\text{regulatn}_i  \right\},\nu \right],
\end{array}
\label{eq:takeBidModel}
\end{eqnarray}
where $\text{bids}_i$ is the number of bids received by each firm $i$, and the explanatory variables are described in the Online Supplement.  For our priors we again consider
\begin{eqnarray}
\begin{array}{cc}
\bbeta \sim N(\bzero,10^5 \bI), & \quad \nu \sim \text{log-Normal}(0,10^5) \ ,
\end{array}
\label{eq:takeBidPrior}
\end{eqnarray}
corresponding to vague {\it a priori} beliefs about the model parameters. Algorithm \ref{alg:compRegBayes} was used to obtain a sequence of random draws from the posterior distribution arising from model \eqref{eq:takeBidModel}--\eqref{eq:takeBidPrior}. Graphical summaries of the MCMC output and subsequent Bayesian inferences for the model parameters are presented in Figure \ref{fig:bidResult}.

The interpretation of the posterior means of each parameter is analogous to those from the first example. For example, firms that employed a legal officer during takeover negotiations (leglrest = 1) are estimated to receive $\exp(0.271) = 1.31$ times as many bids as a comparable firm that did not. The 95\% credible interval for this increase is between $\exp(0)=1$ and $\exp(0.53) = 1.70$ times. The posterior distributions for the other regression parameters can be interpreted in a similar way.
For the dispersion parameter, the posterior mean is 1.62 with a 95\% credible interval of (1.15, 2.06). This indicates that the data exhibit strong underdispersion. 

Again, overlaying the kernel density plots in the fifth column of Figure \ref{fig:bidResult} are the 95\% credible intervals obtained from both the CMP$_\mu$ and standard Poisson models. We see that the credible intervals from the Poisson model are wider than their $\cmpMu$ counterparts across all predictors. In this case, the Poisson distribution has failed to account for the strong underdispersion in these data. The average computation time for each MCMC update here was $0.05$ seconds.

\subsection{Sensitivity to the choice of prior}
\label{sec:sensitivity}
To examine sensitivity of the posterior distributions to the choice of prior, we  re-ran the two data analysis examples using improper flat priors $p(\bbeta ), \ p(\nu) \propto 1$, so that the posterior is directly proportional to the likelihood. These improper priors can be conceptualised as taking the variance hyperparameters in (\ref{eq:meanCmpRegBayes}) to be arbitrarily large. The results from the corresponding MH MCMC are summarised in Figures \ref{fig:daysNoninform} and \ref{fig:bidNoninform} in the Online Supplement. We see that the posterior distributions, means and credible intervals for model parameters are all very similar to the earlier results, suggesting that the proposed framework can be quite robust to the choice of prior, at least for problems that are similar in size and complexity to these two examples.

\subsection{Frequentist coverage rates of credible intervals}
The data analysis examples demonstrate the parsimony of the proposed framework for handling both over and under dispersed counts. Here, we complement these examples by examining frequentist coverage rates of credible intervals for  regression coefficients obtained from the Bayesian CMP$_\mu$ GLM \eqref{eq:meanCmpReg}--\eqref{eq:meanCmpRegBayes}. We compare these coverage rates with those of credible intervals generated from a classical Bayesian Poisson or negative binomial regression model.

Synthetic data are generated from the following three  distributional settings:
\begin{enumerate}
\item no dispersion -- $y_i \simind $ Poisson$(\mu_i$)
\item underdispersion -- $y_i \simind $CMP$_\mu(\mu_i, \nu = 1.62)$
\item overdispersion -- $y_i \simind $ negative binomial($\mu_i,\phi =2$)  with mean $\mu_i$ and variance $\mu_i + \mu_i^2/2$
%\item CMP$_\mu(\mu, \nu = 0.4) $ -- overdispersion
\end{enumerate}
where the means are given by the fitted posterior mean model
\begin{eqnarray*}
\mu_i &=& \exp\big\{0.975 + 0.271 \ \text{leglrest}_i -0.183 \  \text{rearest}_i  + 0.041 \ \text{finrest}_i \\
& & \left.\left. + 0.496 \ \text{whtknght}_i - 0.695\ \text{bidprem}_i - 0.389 \ \text{insthold}_i + 0.183 \ \text{size}_i  \right.\right. \\
& & - 0.008 \ \text{size}^2_i - 0.038\ \text{regulatn}_i  \big\} \ .
\end{eqnarray*}
Each synthetic dataset contains $n=126$ samples, the same as the original takeover bids dataset, and are conditioned on the same set of covariates. For each simulation, the Bayesian CMP$_\mu$ model \eqref{eq:meanCmpReg}--\eqref{eq:meanCmpRegBayes} was fit to the data along with the standard Bayesian Poisson and negative binomial models. For all three models, the prior distribution for the mean parameters was set to $N(\boldsymbol{0}, 10^5\bI)$, whilst the prior distribution for the dispersion parameter was set to Log-Normal$(0,10^5)$ for the CMP$_\mu$ and negative binomial models. Each simulation setting was replicated $N=1000$ times.

The coverage rates of nominal 90\%, 95\% and 99\% credible intervals from all three models are displayed in Table \ref{tab:coverageRateInt}. Note that we have focussed on the coefficients of the first four covariates which correspond to possible defensive actions that could be taken by a target firm. The remaining six coefficients exhibit similar behaviour, but their corresponding covariates are not actionable factors at the firm level and so are not as interesting to look at. 

We see from  Table \ref{tab:coverageRateInt} that for counts with no dispersion the credible intervals from all three models are in agreement with their nominal coverage rates. When counts are generated from an overdispersed negative binomial distribution, credible intervals from both the negative binomial and the misspecified $\cmpMu$ models have coverage rates that remain close to their nominal levels and comparable to the correctly-specified negative binomial model, but the Poisson model undercovers considerably. For underdispersed counts, coverage rates of the negative binomial model are omitted because they cannot handle underdispersion -- in fact, the limiting case would coincide with the Poisson model. Indeed, the Poisson posterior credible intervals are far too wide with the true mean covered at a much higher rate than the nominal levels. In contrast, the $\cmpMu$ model performs well in all three scenarios, reflecting its flexibility in handling equi, over and under dispersion equally well. 

We also examine coverage properties of posterior credible intervals for the dispersion parameter $\nu$ using the proposed Bayesian CMP$_\mu$ approach. In Table \ref{tab:coverageRateInt} we see that for both Poisson and underdispersed CMP data, the coverage rates for the dispersion parameter $\nu$ were again close to their nominal rates. 
For overdispersed negative binomial data, the CMP$_\mu$ model is misspecified -- we can instead look at the {\it power} of the posterior credible intervals, noting that nominal 95\% credible intervals they did {\it not} contain $\nu=1$ in 98.0\% of simulations. For narrower 90\% credible intervals the power improved to 99.1\%, and for wider 99\% credible intervals the power was at 93.4\% in our simulations. These results again suggest that the $\cmpMu$ framework can be equally useful for handling equi, over and under dispersion under both correctly specified and misspecified data-generating mechanisms. Moreover, additional simulations carried out by the second author under different sample sizes and different levels of dispersion yielded similar results to those presented in Table \ref{tab:coverageRateInt}.

\begin{table}[!ht]
%\fontsize{9}{11}\selectfont
\begin{scriptsize}
\begin{center}
\caption{Coverage rates (\%) of nominal 90\%, 95\% and 99\% credible intervals for the first four coefficients $\beta_1, \beta_2, \beta_3$ and $\beta_4$ from the takeover bids example using the Poisson, negative binomial and $\cmpMu$ models under equi, over and under dispersion. $N= 1000$ simulations each with $n=126$ sample size.}
\begin{tabular}{l<{\hspace{-2pt}}l<{\hspace{-2pt}}l<{\hspace{-2pt}}l<{\hspace{-2pt}}l<{\hspace{-2pt}}l<{\hspace{-2pt}}l<{\hspace{-2pt}}l<{\hspace{-2pt}}l<{\hspace{-2pt}}l<{\hspace{-2pt}}l<{\hspace{-2pt}}l<{\hspace{-2pt}}l<{\hspace{-2pt}}
l<{\hspace{-2pt}}l<{\hspace{-2pt}}l<{\hspace{-2pt}}
l<{\hspace{-2pt}}l<{\hspace{-2pt}}l<{\hspace{-2pt}}}
\hline
\hline
& & \multicolumn{17}{c}{Data generating distribution} \\
\hline 
  &  &  &  \multicolumn{5}{c}{Poisson$(\mu_i)$} & &   \multicolumn{4}{c}{Neg-Bin($\mu_i, \phi = 2)$} & &   \multicolumn{5}{c}{CMP$_\mu(\mu_i, \nu =1.76)$}  \\
Model  & level & & $\beta_1$  &  $\beta_2$  & $\beta_3$ &  $\beta_4$ & $\nu $  & & $\beta_1$  &  $\beta_2$  & $\beta_3$ &  $\beta_4$  & & $\beta_1$  &  $\beta_2$  & $\beta_3$ &  $\beta_4$ & $\nu$  \\
\hline
                  &  $90\%$ &  &  87.2 & 86.6 & 85.9 & 88.2 & & & 74.8 & 72.4 & 71.2 & 75.5 & & 94.2 & 92.9 & 93.2 & 93.8 & \  \\ 
Poisson    &  $95\%$ & &  92.1 & 92.1& 92.6 & 93.9 & \ -- & & 81.5 & 79.7 & 78.8 & 82.4 & & 97.1 & 96.9 & 97.5& 96.5 & \ -- \\
                 &  $99\%$ &  & 96.8 & 97.0 & 97.4 & 98.1 & & & 89.4 & 88.8 & 88.0 & 89.8 & & 99.6 & 99.5 & 99.3 & 99.7 & \  \\         
\hline
                & $90\%$ & & 87.5 & 88.0 & 86.9 & 87.5 & & & 85.4 & 84.0 & 83.2 & 85.2 \\ 
Neg-Bin & $95\%$ & &  93.1 & 92.2 & 92.1 & 93.1 & \ -- & & 90.7 & 89.4 & 89.8 & 90.6 & & & \multicolumn{3}{c}{--}  & \\
               & $99\%$ & & 97.2 & 97.1 & 96.4 & 96.9 & & & 95.2 & 95.1 & 95.0 & 95.4 & \\          
\hline
                    & $90\%$ & & 87.4 & 87.5 & 86.3 & 86.5 & 88.1 & & 83.7 & 87.1  & 85.4  & 87.6 & & 91.5 & 90.1 & 89.0 & 87.5 & 88.7 \\ 
$\cmpMu$ & $95\%$ & & 94.5 & 91.9 & 92.5 & 93.9 & 93.4 & & 91.0 & 89.2 & 93.0 & 91.3 & & 95.2 & 93.7 & 92.8 & 95.5 & 93.8 \\
                   & $99\%$ & & 98.5 & 96.0 & 96.5 & 98.5 & 97.8 & & 96.9 & 98.0 & 97.2 & 96.4 & & 97.5 & 98.2 & 97.8 & 98.1 & 98.6 \\          
\hline
\hline
\end{tabular}
\label{tab:coverageRateInt}
\end{center}
\end{scriptsize}
\end{table}

\section{Discussion}

This note proposes a Bayesian generalized linear model framework for both over and under dispersed counts. While it is demonstrated to be particularly simple to implement in practice and exhibits good frequentist coverage properties, we postulate that the proposed approach can be further improved by making it more efficient using the techniques of \cite{CENN2017} and \cite{Benson2017}. This has been earmarked for future research, along with extending the approach to hierarchical mixed models for dependent counts as well as including covariates in the dispersion parameter to account for non-constant dispersion.

\section*{Acknowledgements}
We thank the Associate Editor and two anonymous referees for comments and suggestions that improved this paper. We thank Dr. Thomas Fung (Macquarie University) for help with the \texttt{mpcmp} package in \texttt{R}.

\section*{Supplementary material}
The online supplement contains mathematical details of the conjugate prior for CMP$_\mu$ distributions, a description of the predictors for the takeover bids dataset, and summaries of posterior inferences for the two data analysis examples using flat priors.
\texttt{R} code implementing the proposed algorithm can be obtained by emailing the authors.

\newpage

\section*{\Large Supplementary material for ``Bayesian generalized linear model for over and under dispersed counts"}

\begin{center}
Alan Huang, and Andy Sang Il Kim
\end{center}

\section*{S1. Conjugate prior for CMP$_\mu$ distributions}
$\cmpMu$ distributions form two-parameter exponential families \citep[Appendix 1.2,][]{Huang2017}. We can therefore obtain the conjugate prior and posterior for the $\cmpMu$ distribution by substituting $\lambda=\lambda(\mu,\nu)$ into equation (8) of \cite{KSMBB2006}. Then, we can write the the density function of the conjugate prior for the $\cmpMu$ distribution in the form of 
\begin{eqnarray*}
p(\mu,\nu) &=& \lambda(\mu,\nu)^{a-1} \exp\left(-b\nu \right) Z^{-c}(\lambda(\mu,\nu),\nu)\kappa(a,b,c) \ , \qquad \mu >0 \ , \nu \ge 0 \ ,
\label{eq:conjCmp}
\end{eqnarray*}
where $\kappa(a,b,c)$ is a normalising constant given by
\begin{eqnarray*}
\kappa^{-1}(a,b,c) = \int^{\infty}_0 \int^{\infty}_0  \lambda(\mu,\nu)^{a-1}\exp(-b\nu)Z^{-c}(\lambda(\mu,\nu),\nu) d\mu \ d\nu,
\label{eq:conjCmpNC}
\end{eqnarray*}
and $a$, $b$, and $c$ are hyperparameters restricted by (10) from \cite{KSMBB2006}. The corresponding posterior is of the same form,
\begin{eqnarray*} 
p(\mu,\nu \vert \ \by) &=& \lambda(\mu,\nu)^{a'-1} \exp\left\{-\nu  b'\right\} Z^{-c'}(\lambda(\mu,\nu),\nu)\kappa(a',b',c') 
\label{eq:postConjCmp}
\end{eqnarray*}
with $a'=a+\sum^{n}_{i=1} y_i $, $b'=b+\sum^{n}_{i=1} \log (y_i!)  $ and $c'=c+n$. 

\section*{S2. Description of predictors for takeover bids dataset}
The following description of explanatory variables are taken from \cite{SS2013}:
\begin{itemize}
\item Defensive actions taken by management of target firm: indicator variable for legal defense by lawsuit (\texttt{leglrest}), proposed changes in asset structure (\texttt{rearest}),  proposed change in ownership structure (\texttt{finrest}) and management invitation for friendly third-party bid (\texttt{whtknght}).
\item Firm-specific characteristics: bid price divided by price 14 working days before bid (\texttt{bidprem}), percentage of stock held
by institutions (\texttt{insthold}), total book value of assets in billions of dollars (\texttt{size}) and book value squared (\texttt{size$^2$}).
\item Intervention by federal regulators: an indicator variable for Department of Justice intervention (\texttt{regulatn}).
\end{itemize}
% end of file template.tex

\section*{S3. Class attendance and takeover bids data analysis examples using improper flat priors}
\begin{figure}[]
\centering
\includegraphics[width= 0.8 \textwidth]{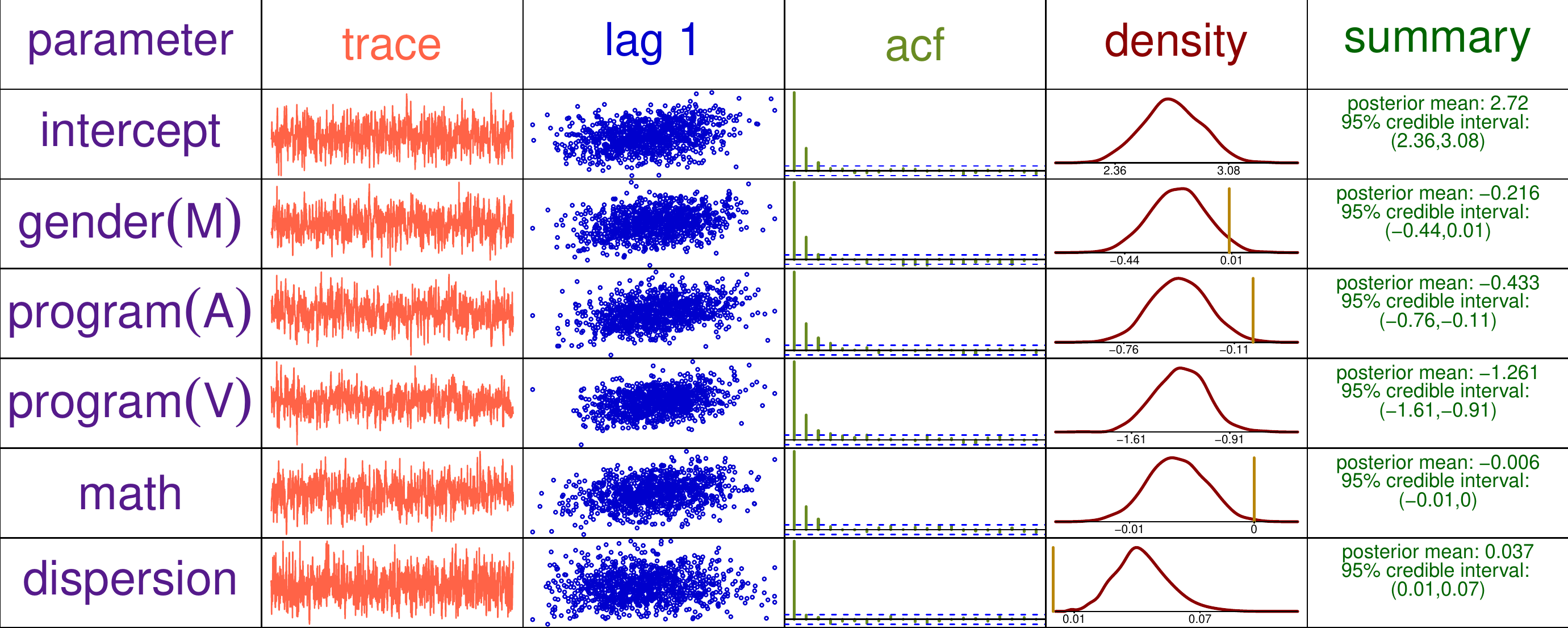}
\caption{Summary of MH MCMC-based inference for parameters in the $\cmpMu$ log-linear Bayesian regression model \eqref{eq:meanCmpRegBayes} fitted to the class attendance data using improper flat priors $p(\bbeta), p(\nu) \propto 1$.}
\label{fig:daysNoninform}
\end{figure}

\begin{figure}[]
\centering
\includegraphics[width= 0.8 \textwidth]{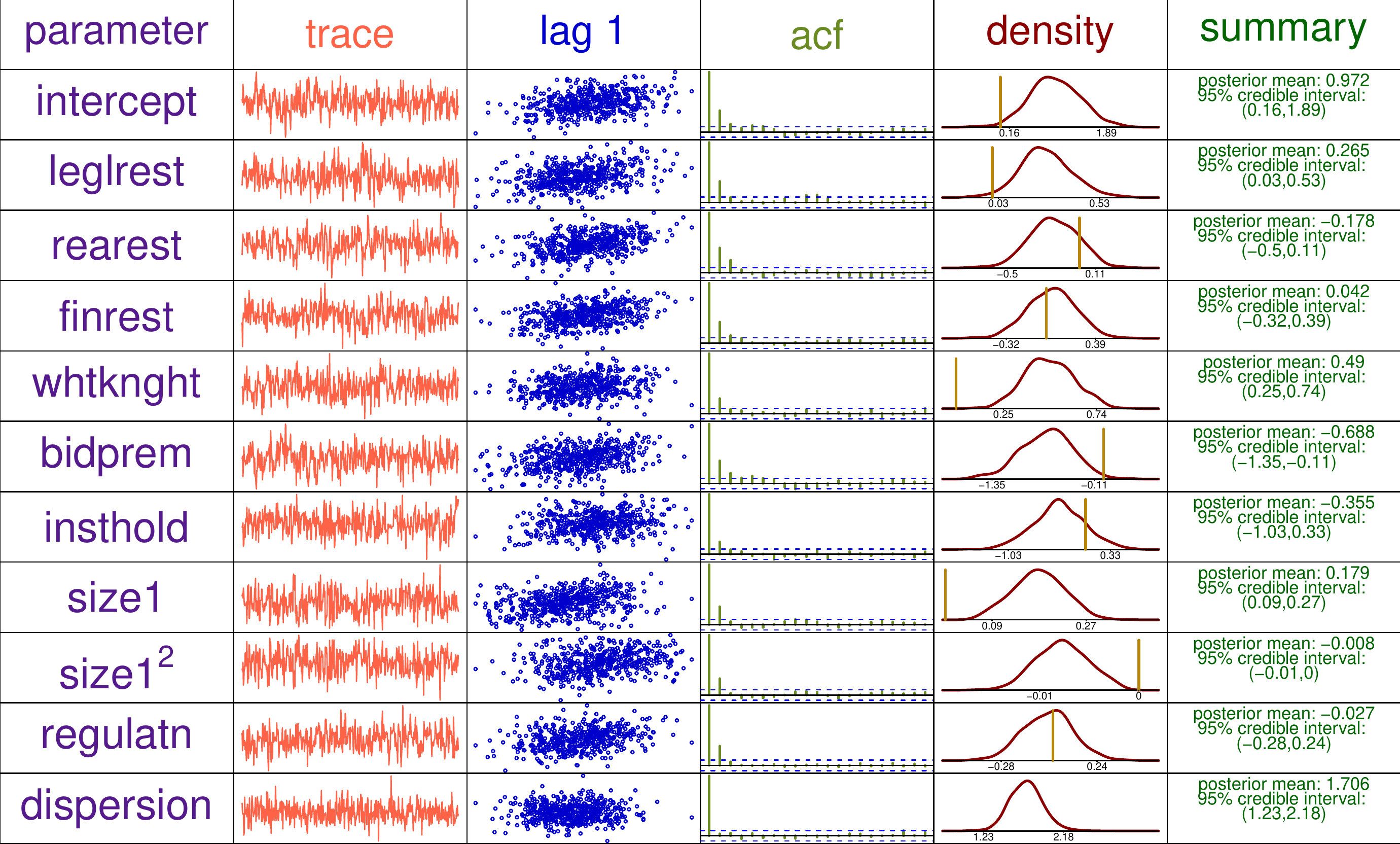}
\caption{Summary of MH MCMC-based inference for parameters in the $\cmpMu$ log-linear Bayesian regression model \eqref{eq:meanCmpRegBayes} fitted to the takeover bids data using improper flat priors $p(\bbeta), p(\nu) \propto 1$.}
\label{fig:bidNoninform}
\end{figure}

The results of the class attendance and takeover bids data analysis examples using the improper flat priors, $p(\bbeta), p(\nu) \propto 1 $, are summarized in Figures \ref{fig:daysNoninform} and \ref{fig:bidNoninform} respectively. These results are very similar to the results based on the Normal and log-Normal pair of priors from the main text, suggesting that the framework is robust to the choice of prior, at least for problems with similar size and complexity to these two examples.

\end{document}